\pdfoutput=1

\documentclass[manuscript]{acmart} 

\usepackage[autosize]{dot2texi}
\usepackage{tikz}
\usepackage{tikzscale}
\usetikzlibrary{shapes,arrows}

\usepackage{caption}
\usepackage{subcaption}

\newdimen\figrasterwd
\figrasterwd\textwidth

\usepackage{cleveref}

\AtBeginDocument{%
  \providecommand\BibTeX{{%
    \normalfont B\kern-0.5em{\scshape i\kern-0.25em b}\kern-0.8em\TeX}}}

\setcopyright{acmcopyright}
\copyrightyear{2022}
\acmYear{2022}
\acmDOI{XXXXXXX.XXXXXXX}

\acmConference[CHI-Play 2022]{Make sure to enter the correct
  conference title from your rights confirmation emai}{November 02--05,
  2022}{Bremen, Germany}
\acmBooktitle{CHI-Play 2022,
 November 02--05, 2022, Bremen, Germany} 
\acmPrice{xx}
\acmISBN{xxx}

\begin{document} 

\title[Towards SAAG in a Multiplayer Environment using AR and Carcassonne]{Towards Situation Awareness and Attention Guidance in a Multiplayer Environment using Augmented Reality and Carcassonne}


\author{David Kadish}
\email{david.kadish@mau.se}
\orcid{0000-0003-3573-2917}
\author{Arezoo Sarkheyli-Hägele}
\email{arezoo.sarkheyli-haegele@mau.se}
\author{Jose Font}
\email{jose.font@mau.se}
\affiliation{%
  \institution{Department of Computer Science and Media Technology, Malmö University}
  \city{Malmö}
  \country{Sweden}
}

\author{Diederick C. Niehorster}
\affiliation{%
  \institution{Lund University Humanities Lab and Department of Psychology, Lund University}
  \city{Lund}
  \country{Sweden}
  }
\email{diederick_c.niehorster@humlab.lu.se}

\author{Thomas Pederson}
\affiliation{%
  \institution{University West}
  \city{Trollhättan}
  \country{Sweden}
  }
\email{thomas.pederson@hv.se}

\renewcommand{\shortauthors}{Kadish, et al.}

\begin{abstract}
  Augmented reality (AR) games are a rich environment for researching and testing computational systems that provide subtle user guidance and training.
  In particular computer systems that aim to augment a user's situation awareness benefit from the range of sensors and computing power available in AR headsets.
  In this work-in-progress paper, we present a new environment for research into situation awareness and attention guidance (SAAG): an augmented reality version of the board game Carcassonne.
  We also present our initial work in producing a SAAG pipeline, including the creation of game state encodings, the development and training of a gameplay AI, and the design of situation modelling and gaze tracking systems.
\end{abstract}



\maketitle

\section{Introduction}

Augmented Reality (AR) enhances the real world with the addition of computer generated virtual elements. Many senses, smell, touch, hearing, and sight, can potentially be augmented, though the most common application of AR is sight, using a head-mounted display~\cite{carmigniani2011augmentedsurvey}. Several users may simultaneously access and operate a shared digitally augmented environment, either at the same place or remotely. Users commonly interact with each other and the augmented elements in this virtual framework by using hand gestures, movement, and even gaze. The interactive nature of AR, as well as its direct connection to the real world, have produced extensive research work and industrial applications of AR to different fields such as education, entertainment, medicine, and retail \cite{huang2019augmentedvsvirtual}.

Human-Computer interaction in games (HCI-games) is a very broad field that covers research on the many ways in which human players interact with digital games that, given their interactive, playful, and challenging nature, present a rich field of study separated from human-computer interaction in other forms of software \cite{BARR2007180hcigames}. This also makes digital games a very suitable environment for other areas of research, such as artificial intelligence, user-centered design, situation awareness, pervasive interfaces, collaborative and competitive behavior, and long-term planning \cite{yannakakis2018artificial,hinske2007classifying}.


This paper proposes an AR version of the acclaimed tabletop game Carcassonne as a framework for developing a combined situation awareness and attention guidance system for multi-user digital environments.
Here we present the game which we refer to as cARcassonne, the framework for a situation awareness and attention guidance (SAAG) system that we have built using cARcassonne as a testbed, and the initial design of a number of the SAAG modules.
%

\section{Related Work}

\subsection{Augmented Reality and Games}

The rise and evolution of AR technologies is closely related to the advancements in the hardware used for its implementation;
this includes displays such as headsets combined with a variety of sensors and controllers that enable the user to interact in a mixed reality environment either virtual or physical objects.
The applications of AR have been many in areas such as retail, healthcare, immersive prototyping, education, aeronautic, and military \cite{parekh2020systematic}.


One of the most prolific sectors in the application of AR and mixed-reality has been games.
As in other areas, games are a very fitting platform for prototyping and benchmarking approaches that make use of relatively unexplored technologies or novel algorithms \cite{yannakakis2014panorama}.
The immersive and world-integrated nature of AR games provides a unique opportunity for studying computer augmentation and guidance of real-world tasks.
In addition to creating the ability to add audiovisual cues to a user's environment, AR headsets include sensor systems that can track a user's response to stimulus, including motion sensing and gaze tracking.
These digital sensors can feedback into the process of learning from user behaviour to further improve the augmentation systems.

\subsection{Situation Awareness}

Among the fields that can be advanced by AR is \textit{situation awareness}~\cite{Endsley1995}.
Situation awareness, originally defined as "the perception of the elements in the environment within a volume of time and space, the comprehension of their meaning, and the projection of their status in the near future"~\cite{Endsley1995}, is often applied to complex, high-stress situations such as the processing of information by an aircrew in the cockpit of an airliner.

However, AR games provide a low-risk environment in which a user's situational awareness can be assessed and computational tools to improve their awareness can be prototyped and tested.
Ensley's original situation awareness model proposes 3 phases or levels of situational awareness: \textit{perception} of the situation, \textit{comprehension} of the perceived data, and \textit{projection} of the situation into the future~\cite{Endsley1995}.
AR systems have a wealth of tools that can be used to perceive the situation in analog and digital spaces.
They also have the computing power to process those inputs into models of the situation that can also be used to project future situations.

\section{System overview}

The ultimate goal of this research is the production of a computational system to augment a user's situational awareness during Carcassonne gameplay.
The intent to be able to assist a novice player in appreciating the full extent of a scenario by identifying their intended approach and gently guiding them towards more advanced strategies.

\begin{figure}
    \centering
    \includegraphics[width=0.65\textwidth]{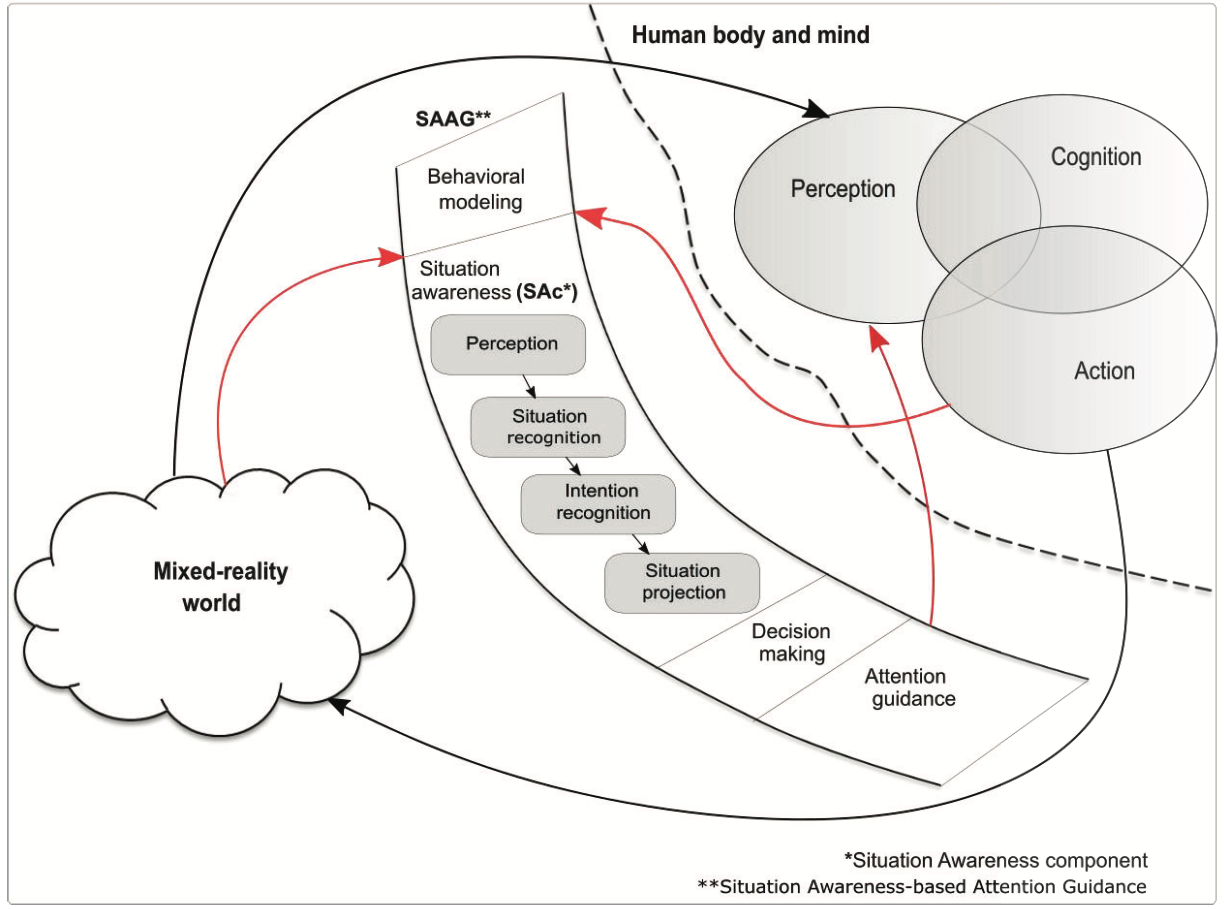}
    \caption{A model of the Situation Awareness and Attention Guidance system.}
    \label{fig:saag}
\end{figure}

A visual representation of this approach is shown in \cref{fig:saag}.
The approach is based on \cite{Endsley1995}, but updates the original model of situation awareness to differentiate between computational processes and human processing of the situation.
The work presented in this paper focuses on the first 3 components of the situational awareness component (SAc) of SAAG: Perception, Situation recognition, and Intention recognition.

Perception refers to how the computer is able to gather information about the system state.
As the gameplay is digital, there is no sensory perception performed by the system per se, but the encoding of game data is part of the perception process.
A pair of encodings of the Carcassone game state, detailed in \cref{sec:encoding} are a major contribution of this work and represent the perception block in the SAAG process.
Situation recognition occurs in the form of the situation model described in \cref{sec:situation}.
\textit{Situation} is defined, in this work, as the likelihood of each of the legal placements of the next tile in play in a given game state and the situation model attempts to capture the typical decisions that a player might make in a particular scenario.
Intention recognition utilizes the data representing a user's gaze along with the situation model to attempt to discern their likely placement.
This module is in its infancy and its initial development is discussed in \cref{sec:gaze}, while planned future work is detailed in \cref{sec:future}.

One challenge that arose in the development of these systems was the lack of available datasets detailing the progression of Carcassonne games.
Our solution was to develop an AI to play Carcassonne, both to generate large databases of AI-versus-AI games that could be used in training the situation model, and to provide an AI opponent for human players.
This AI is detailed in \cref{sec:gamplayai}.

%


\subsection{Carcassonne and cARcassonne}
Carcassonne \cite{carcassone} is is a tile-based strategic board game for two to five players, originally published in 2000, and awarded with the Spiel des Jahres prize in 2001. This preceded an enormous popularity in the coming years that has made it, and its countless expansions and updates, one of the best-selling tabletop games in history. 

It is named after the medieval fortified town of Carcassonne in southern France. This setting is used to challenge players to build, tile by tile, a common landscape where features such as cities, roads, and fields can be created, owned by players, and then scored to win the game. In their turn, players draw a random tile from a common stack, that they then have to place adjacent to any other already placed tile on the board, in a way that their composing features adequately aligned (i.e. roads and cities don't end abruptly and build onto each other). Finally, players may place a so called meeple on one of the features in the placed tile to be able to obtain points from it, provided that it is not previously owned by other player. However, it is possible for some features to be shared by opposing players, either intentionally or not, opening scope for punctual player-player interaction in both, competitive and cooperative fashions.

\begin{figure}
     \centering
     \begin{subfigure}[b]{0.25\textwidth}
         \centering
         \includegraphics[width=\textwidth]{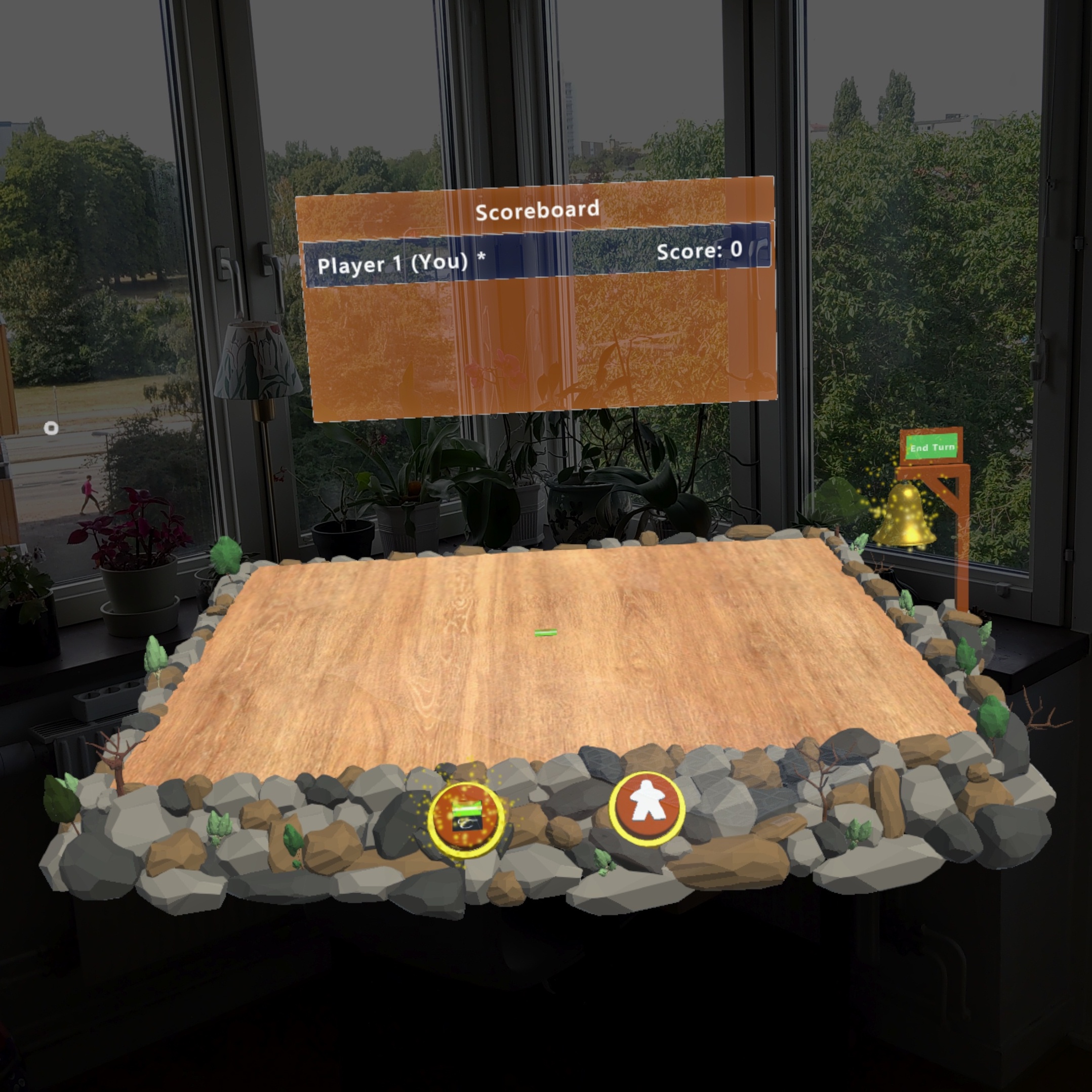}
         \caption{Player's overview of the game}
         \label{fig:mrc_overview}
     \end{subfigure}
     \hfill
     \begin{subfigure}[b]{0.25\textwidth}
         \centering
         \includegraphics[width=\textwidth]{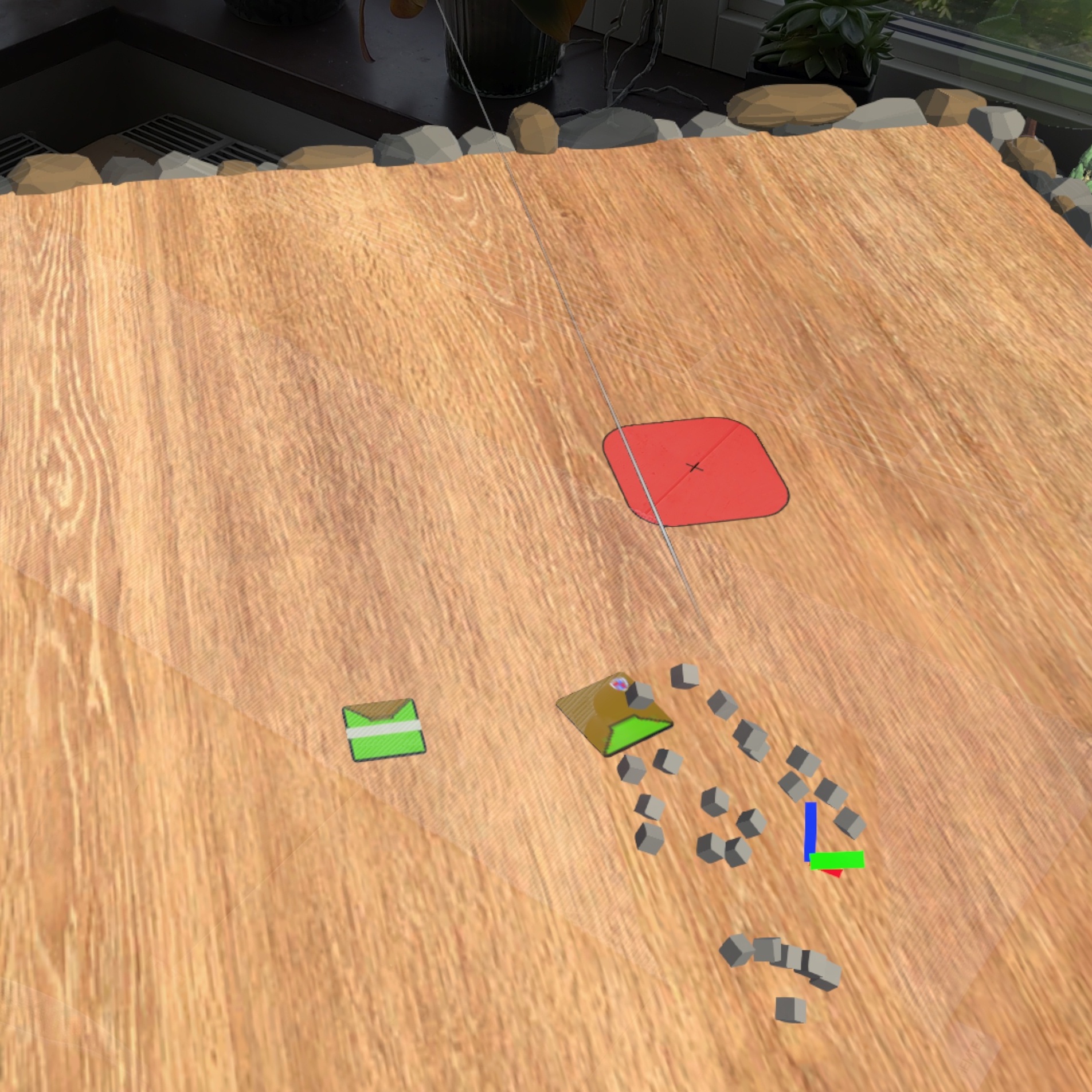}
         \caption{Placing a tile}
         \label{fig:mrc_placing}
     \end{subfigure}
     \hfill
     \begin{subfigure}[b]{0.25\textwidth}
         \centering
         \includegraphics[width=\textwidth]{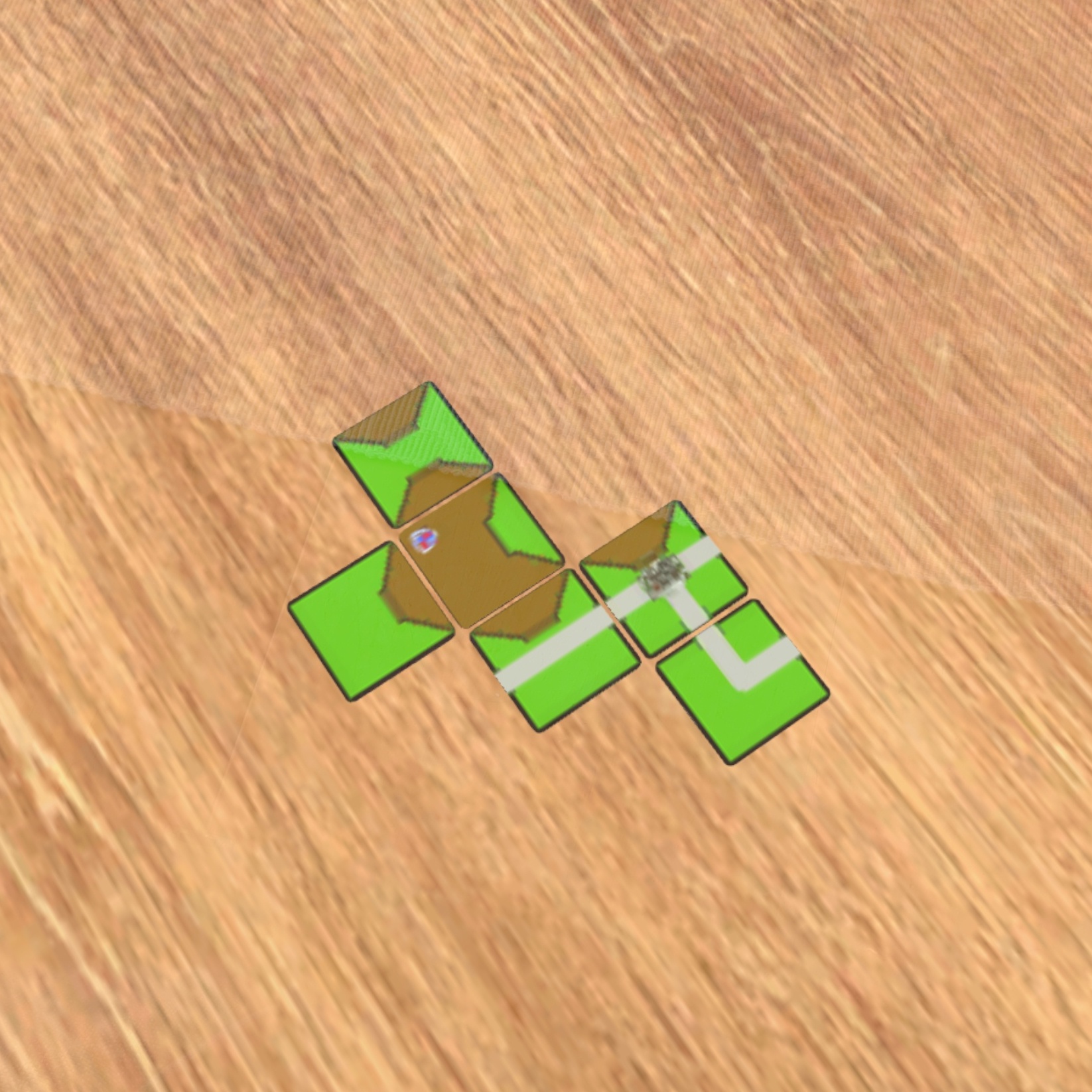}
         \caption{Close-up view of board}
         \label{fig:mrc_close}
     \end{subfigure}
        \caption{Views of cARcassonne game space, as seen by a Hololens 2 user.}
        \label{fig:mixed_reality_views}
\end{figure}

The AR version of Carcassonne, cARcassonne, that we have been developing for this research implements the most basic ruleset of the original board game.
It is designed to he played using the Hololens 2 AR headset, which projects a virtual table into the player's physical environment (\cref{fig:mixed_reality_views}).
Players draw tiles and meeples using a virtual button and then are able to pick up, rotate, and place the game pieces with actions that are similar to playing a physical version of the board game.
Audiovisual cues indicate to the player whether their placement is legal and their turn is ended by ringing a virtual bell that sits at the side of the table.

\subsection{Game state encoding}
\label{sec:encoding}

The representation of a game that is used in computational processing informs the way that the computational systems are able to learn models and representations of the game situation.
It is important that the game state encoding is able to capture all important features of the game in a compact and efficient representation.

We developed two complimentary game encodings for the Carcassonne board that are used in different contexts in the game and the SAAG pipeline.
A bit encoding (\cref{sec:bit}) is used in visualizations of the game board as well as in the convolutional neural network (CNN)-based learning processes that train the gameplay AI (see \cref{sec:gamplayai}).
A graph encoding (\cref{sec:graph}) is used heavily in the game engine to determine legal moves and scoring of game features, but also forms the basis of the graph neural network model for the situation model (see \cref{sec:situation}).

\begin{figure}
     \centering
     \begin{subfigure}[b]{0.2\textwidth}
         \centering
         \includegraphics[width=\textwidth]{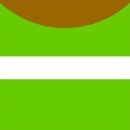}
         \caption{Starting game tile}
         \label{fig:tile}
     \end{subfigure}
     \hfill
     \begin{subfigure}[b]{0.2\textwidth}
         \centering
         \includegraphics[width=\textwidth]{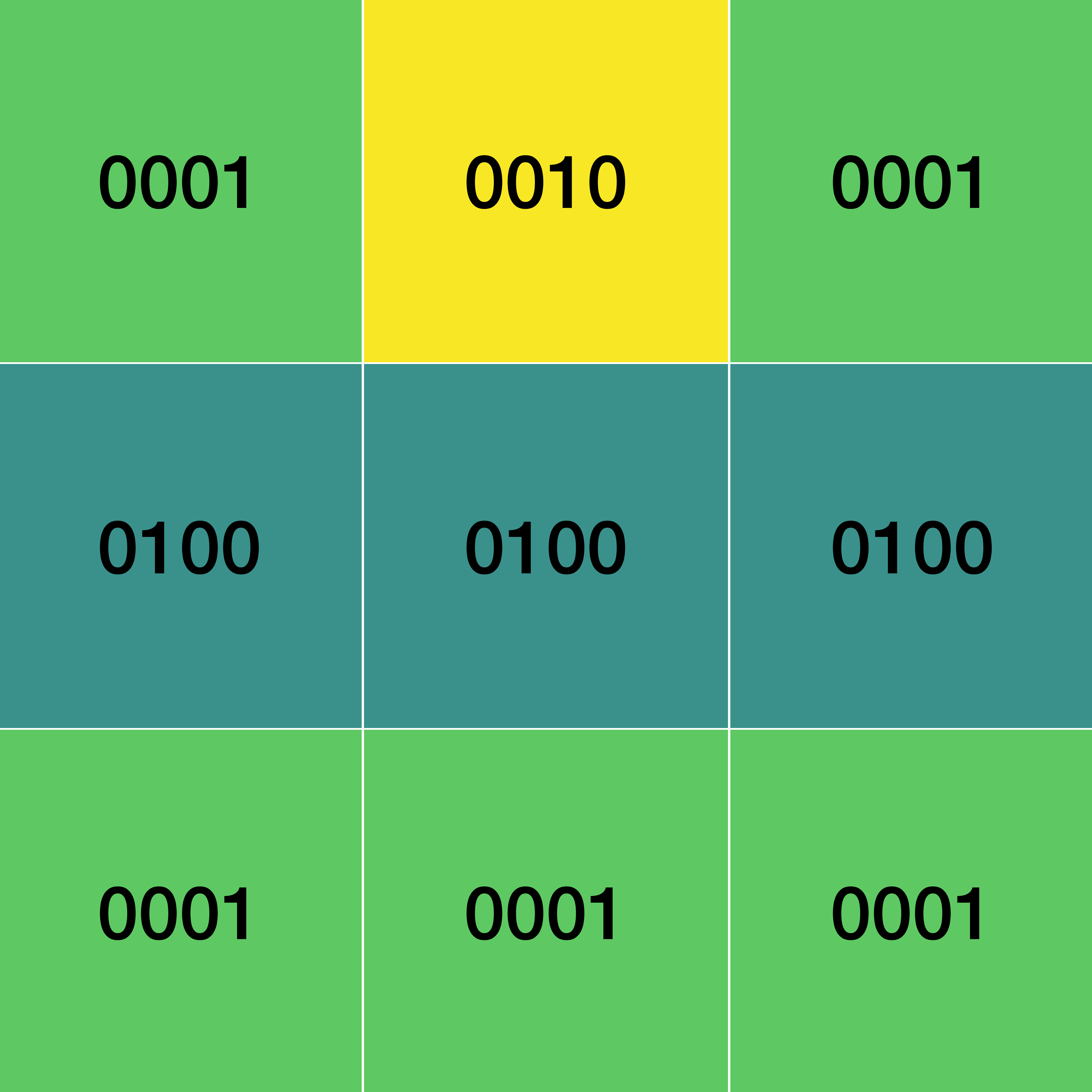}
        \caption{Bit encoding}
        \label{fig:bit}
     \end{subfigure}
     \hfill
     \begin{subfigure}[b]{0.2\textwidth}
        \centering
         \includegraphics[page=2,width=\textwidth]{images/graph_encoding.pdf}
        \caption{Graph representation}
        \label{fig:graph}
     \end{subfigure}
    \caption{A single game tile (the starting tile) with representations of its two encodings. The tile features a city at the top and a road that runs through the middle with one field between the city and road and another running along the bottom.}
    \label{fig:representations}
\end{figure}

\subsubsection{Bit encoding}
\label{sec:bit}
In this representation, each tile is divided into a 3x3 sub-tile grid.
Each grid cell is represented by a bit field that holds a representation of the features in that part of the tile including for cloisters, roads, cities, and fields.
The entire game board is represented as a $120 \times 120$ sub-tile matrix\footnote{We found experimentally that games do not extend more than 16 tiles in any direction from the initial tile, so that a $40\times 40$ game board is sufficient to capture the extent of normal gameplay. Each tile has a $3\times 3$ sub-tile representation, so a $120 \times 120$ matrix can represent the whole board as sub-tiles.} where spaces without placed tiles represented by $0000$.
This representation forms the basis for the tensors generated for gameplay AI training in \cref{sec:gamplayai}.

\subsubsection{Graph encoding}
\label{sec:graph}

Carcassonne is a game of connections, so it lends itself well to a graph-based representation.
Therefore, we chose to represent the state of the game board as a graph for many of the key tasks in this work.

Each Carcassonne tile consists of four potential connection points (graph vertices) --- one on each side of the square tile.
The tile may also contain a vertex in the middle, if a cloister is present.
Three types of edges are used to connect the vertices: intra-tile, inter-tile, and feature.
Intra-tile connections (purple) are permanent features of each game tile and connect all physically adjacent vertices.
They represent the physical, spatial layout of the vertices on a tile.
Feature connections (yellow) represent the logical connections between vertices.
Vertices on a single road or that form a single city are all connected by a feature connection.
Inter-tile connections (blue) are formed when two vertices are placed adjacent one another on the game board.

The graph representation is used extensively in the game's underlying logic system.
Graph methods enable rapid and robust computation of feature completion, the legality of game piece placement, and scoring.
This representation also forms the basis for the graph neural network-based situation model described in \cref{sec:situation}.

\subsection{Gameplay AI}
\label{sec:gamplayai}

The gameplay AI is a CNN-based agent that plays Carcassonne, selecting the placement of a tile and an optional meeple based on the current state of the game.
Unity's built-in ML-Agents framework~\citep{Juliani2018} is used to train the agents and to play games once training is complete.
The gameplay AI is used within the SAAG pipeline both to generate game situations for training the situation model (\cref{sec:situation}) and as an opponent for human players.

\subsubsection{Observations}

To engage with the game, the agent receives an encoded representation of the game state, referred to in ML-Agents as \textit{observations}.
Observations are collected as a 120x120x5 tensor with the first two dimensions representing the width and height of the sub-tile matrix of a 40x40 game board.
The final dimension holds five observable variables for each sub-tile: \textit{cloister}, \textit{road}, \textit{city}, \textit{shield}, and \textit{meeple}.
The cloister, road, and city dimensions are drawn from the bit encoding (\cref{sec:bit}) of the game state and these are treated as boolean fields where 0 and 1 indicate the absence and presence of a feature, respectively.
For the \textit{meeple} layer, 0 indicates the absence of a meeple, and placed meeples are indicated with a value representing the player to which the meeple belongs.
In addition, observations indicating the score, current tile state, and number of tiles and meeples remaining are passed to the AI.

The tensor representation is not the default mode of representing data in ML-Agents --- the most basic mode of representation is an observation vector.
However, it is essential here to maintain the spatial relationships between game pieces for the neural network.

\subsubsection{Actions}

Given a set of observations, the neural network produces an action within a defined action space.
To properly limit the set of actions to the legal positions of tiles and meeples on the board, the action space is defined as a vector representing all possible tile rotations, placements, and meeple placements on the game board.
For the $40 \times 40$ game board used here, this means there are 38400 possible actions.
The action space is masked so that, on any given turn, only legal action choices are available.

\subsubsection{Training}
Our initial approach to training used ML-Agents' adversarial self-play mode to train the neural network playing against itself.
Self-play is a form of reinforcement learning where agents play against past iterations of themselves to learn how to play against opponents of ever-increasing difficulty.

The first attempts at this approach were unsuccessful.
Despite training for 1M steps (\~14k games), the agents were not learning even the basic strategies for success in Carcassonne.
They would play all of their meeples immediately instead of reserving some for high-scoring opportunities later in the game, and their tile placements were not designed to expand existing features that they owned.
We next attempted to train the agent in a single-player game with the goal of gaining as many points as possible to see if the types of expected strategies would develop.
Under this system, the agents learned to complete features and we observed the average number of meeples remaining every turn rise.

\begin{figure}
    \centering
     \begin{subfigure}[b]{0.3\textwidth}
         \centering
         \includegraphics[page=2,width=\textwidth]{images/tensorboard/completed_cities}
         \caption{Number of cities completed over the course of a game}
         \label{fig:completed_cities}
     \end{subfigure}
     \hfill
     \begin{subfigure}[b]{0.3\textwidth}
         \centering
         \includegraphics[width=\textwidth]{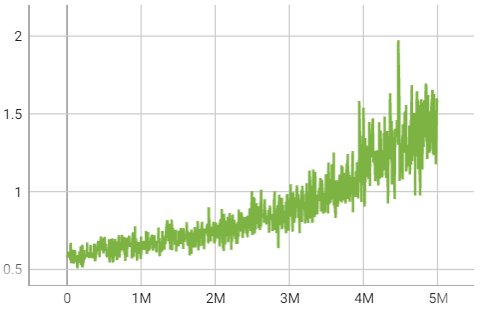}
         \caption{Average number of meeples remaining per turn}
         \label{fig:meeples_remaining}
     \end{subfigure}
     \hfill
     \begin{subfigure}[b]{0.3\textwidth}
         \centering
         \includegraphics[page=3,width=\textwidth]{images/tensorboard/turns_with_point_gain}
         \caption{Number of turns where points were gained}
         \label{fig:point_gain_turns}
     \end{subfigure}
    \caption{Metrics from single-player training. The agents can be seen learning the game strategy as they slowly increase the number of cities that they complete over the course of a typical game. They can also be seen holding on to meeples for longer, and getting meeples back as features are completed, leading to a growth in the average number of meeples remaining every turn. Finally, they are gaining points on more turns, indicating that they are adding to features that they already own instead of randomly placing tiles.}
    \label{fig:training}
\end{figure}

Future work on this module will include the addition of curriculum learning to develop agents that first learn gameplay strategies and subsequently focus on an opponent.
Additionally, human opponents (\cref{sec:future}) will be used to generate datasets for imitation learning that may be able to help learn Carcassonne strategies more effectively.

\subsection{Situation model}
\label{sec:situation}

The situation model is an attempt to model the upcoming moves and flow of the game.
At present, the first attempt at doing so consists of a graph neural network (GNN)~\citep{Kipf2016} that is trained to predict the placement of the next tile given the current board and the tile that has been selected.
The network performs a node-based regression, where it estimates the likelihood of different placements for the upcoming tile.

\subsubsection{Dataset}

\begin{figure}[ht]
  \centering
  \parbox{\textwidth}{
    \parbox{.35\textwidth}{%
      \subcaptionbox{Current board}{\includegraphics[width=\hsize]{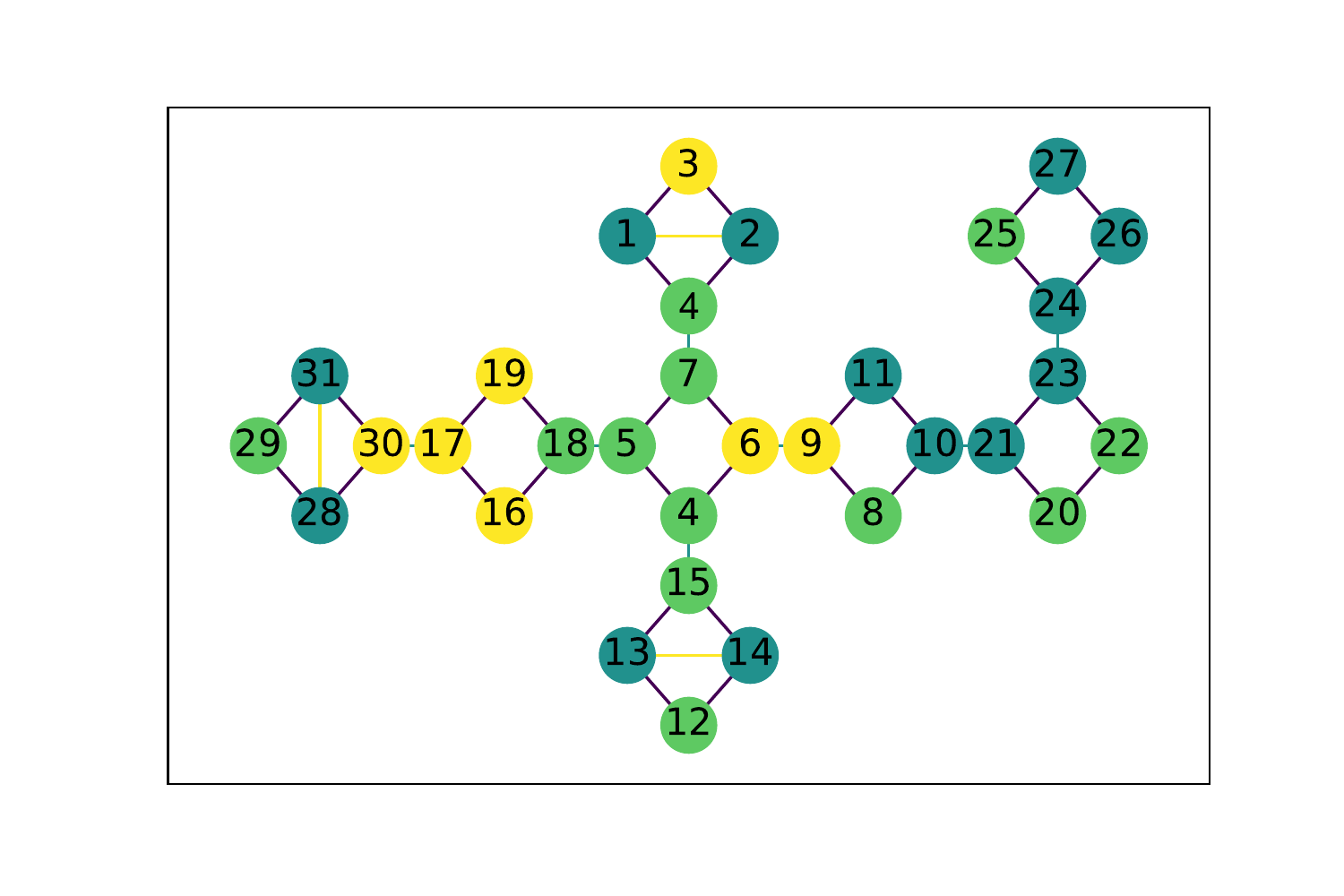}}
      \vskip.25em
      \subcaptionbox{Next tile}{\includegraphics[width=\hsize]{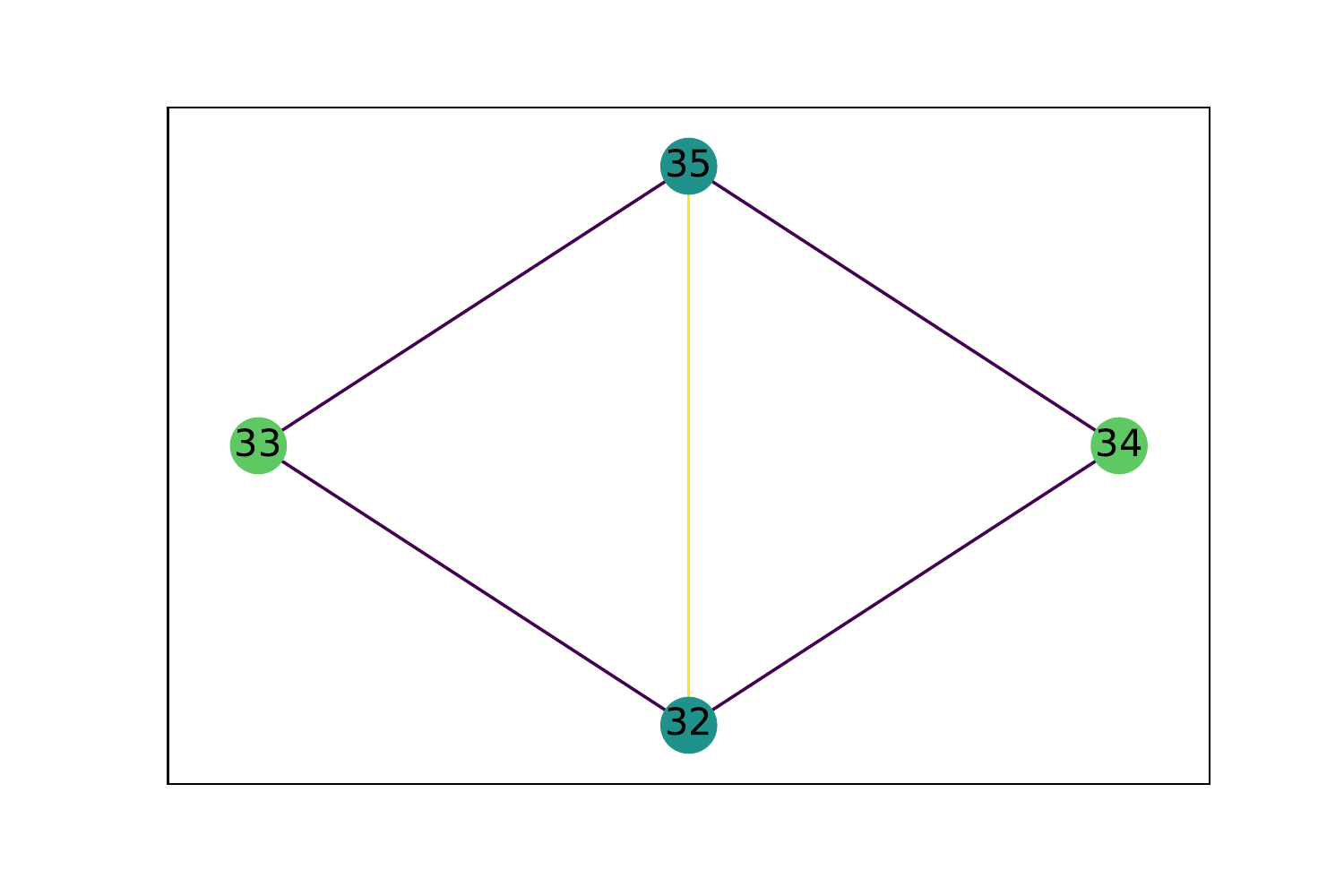}}  
    }
    \parbox{.65\textwidth}{%
      \subcaptionbox{Candidate board (candidates marked with *)}{\includegraphics[width=\hsize]{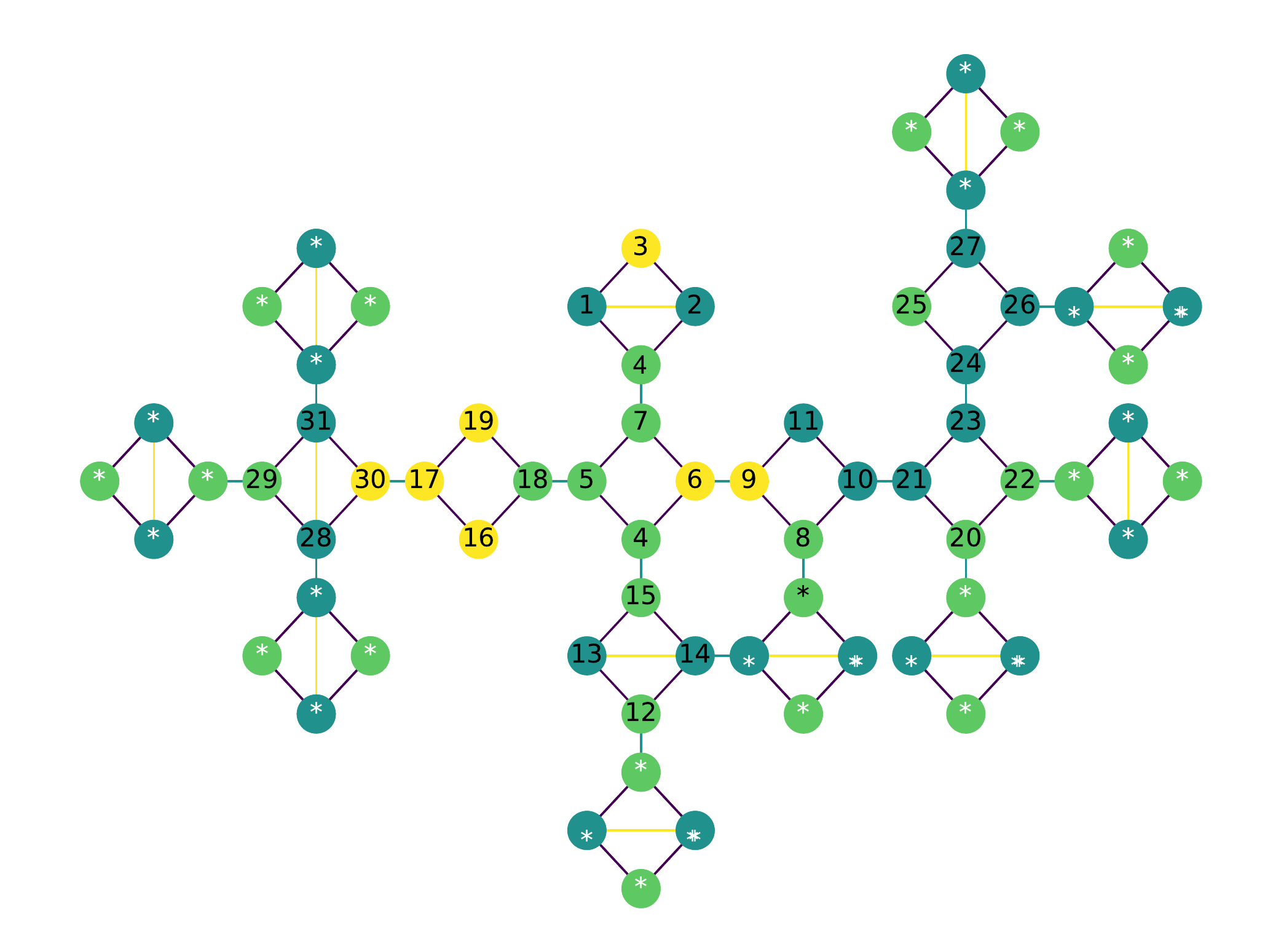}}
    }
  }
  \caption{Graph representations of the current board (a), next tile (b), and the candidate board (c).}
  \label{fig:candidate-graph}
\end{figure}

The situation model is trained on the set of games played by the AI during its training process.
The completed game graph from each playthrough is processed in reverse to generate a set of ~71 board states and next tiles played over the course of that game.
From the combination of the current board graph and the upcoming tile's graph representation, a \textit{candidate board} graph is generated that includes nodes and connections for all legal placements of the incoming tile, with a special attribute set on the nodes indicating that they represent the possible legal moves (\cref{fig:candidate-graph}).

The dataset that is ultimately used to train the neural network consists of this candidate board for each turn along with a ground truth value that indicates which of the candidate tile placements was actually played in the game.

\subsubsection{Model}

The situation model is a graph neural network using the graph convolutional operator from \citep{Kipf2016}.
It consists of two graph convolutional network (GCN)~\citep{Kipf2016} layers separated by a ReLu and dropout layer.
A pooling layer after the final GCN averages the values of all of the nodes on a single tile.
The GNN outputs a floating point number representing the network's estimate of the probability that a given position and rotation will be selected for the upcoming tile.


\subsection{Gaze tracking and data visualization}
\label{sec:gaze}

\begin{figure}
    \centering
     \begin{subfigure}[b]{0.3\textwidth}
         \centering
         \includegraphics[page=2,width=\textwidth]{images/BoardRepresentations}
         \caption{AR game board with graph representation overlay.}
         \label{fig:board_with_graph}
     \end{subfigure}
     \hfill
     \begin{subfigure}[b]{0.3\textwidth}
         \centering
         \includegraphics[width=\textwidth]{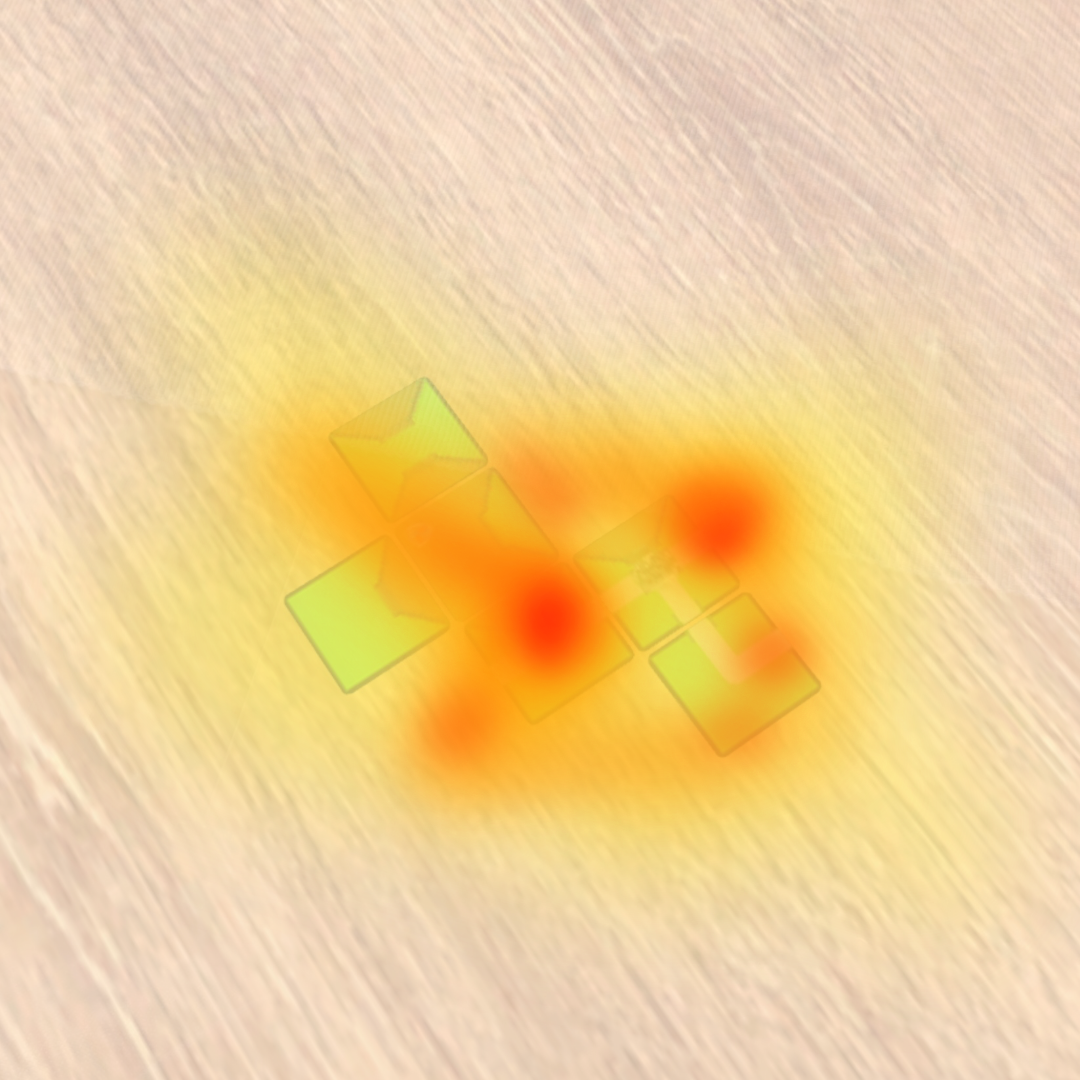}
         \caption{Heatmap showing player gaze, darker colours indicating longer dwell.}
         \label{fig:heatmap}
     \end{subfigure}
     \hfill
     \begin{subfigure}[b]{0.3\textwidth}
         \centering
         \includegraphics[page=3,width=\textwidth]{images/BoardRepresentations}
         \caption{Gaze plot overlayed on board and graph, showing gaze order and duration.}
         \label{fig:gaze_plot}
     \end{subfigure}
    \caption{Mockups of gaze data. A heatmap (b) shows the duration of a user's gaze on different parts of the game board, while the gaze plot (c) captures the order in which a user examines different game board features as well as the duration of the user's gaze at different points. The gaze plot (c) is shown overlayed on the graph representation of the game board (a) as the gaze plot can also be processed as a graph with nodes representing gaze at a particular position and edges representing travel between gaze points.}
    \label{fig:gaze_representations}
\end{figure}

Using the built-in gaze tracking capabilities of the Hololens 2, we have begun work on a module to integrate a measure of the user's visual attention into the SAAG model.
The current implementation logs the user's gaze over a specified period of time and generates a heatmap indicating where they have spent the most time looking (\cref{fig:heatmap}).
Future work will integrate this heatmap into the SAAG pipeline as a partial model of the user's intention.

\subsubsection{Gaze plots as graphs}

Another avenue for the integration of gaze data is also being explored for future implementation in the SAAG pipeline.
Gaze plots record and visualize information not captured by a heatmap including the order and duration of focal points in a user's gaze path.
This temporal aspect of a user's gaze could be important in predicting their intended actions~\citep{Fuchs2021}.
The gaze plot can be treated as a graph with the focal point nodes linked to nodes already present in the situation model, meaning this type of data could be readily integrated in the existing GNN-based model.

\section{Future work}
\label{sec:future}

The implementation and initial testing of the various portions of the SAAG pipeline presented here form the basis for an integrated computational system for augmenting user situational awareness.
However, many of the components suffer from a common problem in machine learning --- the lack of high quality datasets for training\citep{NEURIPS2021_d5eca8dc}.

\subsection{Human players}

In particular, generating realistic human play in the MLAgents framework have proved challenging.
Collecting gameplay data from human players of all levels, however, will provide an improved initial dataset for training the situation model.
It could also serve as basis for a more effective gameplay AI training process, using imitation learning techniques~\citep{Rubak} to develop basic gameplay strategies.
We have planned to conduct a study in collaboration with Lund University's Humanities Lab.
Participants will play a version of the game on a desktop computer in the lab environment against a mixture of human and AI opponents and their games will be recorded for future use in model training.

\subsection{Gaze tracking}

In addition to gameplay data, the Humanities Lab study will also provide an initial set of eye tracking data associated with Carcassonne gameplay.
The Humanities Lab is equipped with Tobii Pro Spectrum gaze trackers~\citep{Nystrom2021} which will enable the capture of high-quality gaze data during gameplay.
This data will be linked to game state data as well as player move choice information.
It will assist in the study of the relationship between gaze at various game phases and the player's strategy and intentions with respect to tile placement.




\begin{acks}
This research is partially funded by the Crafoord Foundation through the project Situation Awareness-based Attention Guidance.
We would like to acknowledge the AR interaction design work by Otto Karlsson Hellström and Kevin Magnusson and the initial ML-Agents integration done by Kasper Skott, Niklas Hultin, Edvin Brus, and Johan Helgstrand.
\end{acks}

\bibliographystyle{ACM-Reference-Format}
\bibliography{saag}

\end{document}